# Pressure-Induced Insulator-to-Metal Transition Provides Evidence for Negative-*U* Centers in Large-Gap Disordered Insulators

Yang Lu and I-Wei Chen*

*Department of Materials Science and Engineering*
*University of Pennsylvania, Philadelphia, PA 19104-6272, USA*
*Correspondence to: iweichen@seas.upenn.edu

**Abstract**: Attractive negative-*U* interactions between electrons facilitated by strong electron-phonon interaction are common in highly polarizable and disordered materials such as amorphous chalcogenides, but there is no direct evidence for them in large-band-gap insulators. Here we report how such negative-*U* centers are responsible for widespread insulator-to-metal transitions in amorphous $HfO_2$ and $Al_2O_3$ thin films with a $10^9$-fold resistance drop. Triggered by a static hydraulic pressure or a 0.1 ps impulse of magnetic pressure, the transition can proceed at such low pressure that there is very little overall deformation (strain~$10^{-5}$). Absent a significant energy change overall, the transition is attributed to the reversal of localized electron-phonon interaction: By reversing the sign of *U*, trapped electrons are destabilized and released, thus clearing conduction paths previously blocked by charged traps. The results also suggest that Mott insulators when disordered may become Anderson insulators with strong electron-phonon interactions regulating incipient conduction paths, a novel finding of technological significance for electronic devices.

**Significance Statement.** *Electron-phonon interactions are responsible for superconductivity, colossal magnetoresistance, Peierl instability and polaron conduction. But its macroscopic demonstration calls for an immediate response in the electromagnetic properties to a mechanical or deformational stimulus. Such demonstration is given here for the first time using a modest* *pressure that spans a wide time scale, to as fast as the period for one atomic vibration—the limiting time scale for a phonon. The experiments further manifest that a Mott insulator, in which strong inter-electron repulsion prohibits electrons from moving around, becomes an Anderson insulator when randomized, and just by chance the disordered network may contain enough sites to host strongly localized electron-phonon interaction and to channel short-distance electron conduction. Nanomaterials with such characteristics may be used for electronic devices.*

Electronic centers usually repel electrons because of electron-electron interactions, among them the positive Hubbard interaction $U$ [1] between opposite-spin electrons in the same local state has a strong influence on electronic properties. However, a negative effective $U$ can also be realized when a strong electron-phonon interaction intervenes [2-5]. Anderson illustrated how a negative $U$ is linked to electron-phonon interaction by the following example: When a second electron is added to a bonding orbital, the bond retracts; likewise, when a second electron is added to an antibonding orbital, the bond extends [2]. Therefore, the freedom to adjust bond length thus stabilizing the bonding orbital is crucial for attracting an extra electron. Since strong electron-phonon interaction and bond distortions require a high polarizability and disorder in the solid state [6], abundant negative $U$ centers were first seen in amorphous chalcogenides where they open up a gap of the order of $U$ [7]. Negative-$U$ centers have also been described for crystalline insulators such as quartz and cubic hafnia, at point defects and surfaces [8-11], and such attractive centers are thought to increase the leakage current in amorphous $SiO_2/HfO_2$ films causing insulation degradation [9,10]. But direct evidence for their existence is scant because these oxides have either wide band gaps well in excess of $|U|$ or quite populous paramagnetic centers (see, e.g., EPR studies of amorphous $SiO_2$ [12] and $HfO_2$ [13,14]) that will mask any diamagnetic signature from electron pairing at negative $U$ centers.

Unlike their crystalline counterpart, amorphous insulators always suffer bond-length/angle dispersions and configurational frustrations. So they invariably contain some local atomic arrangements that can relatively easily relax without straining the surrounding very much—thus without much elastic-energy penalty [15,16]. These "soft spots", just like a rattling $Ti^{4+}$ in a $BaTiO_3$ crystal, should be highly polarizable even though the bulk material, on average, is not. So they could offer negative-$U$ centers to host self-trapped electrons. If the amorphous structure also contains some nanoscale paths that should have been conducting were they not blocked by these negatively charged, self-trapped electrons, then detrapping could dramatically lower its nanoscale resistance. Below we will report how we literally "squeezed" pristine amorphous $HfO_2$ and $Al_2O_3$ to "pinch off" their negative-$U$ centers, and by doing so dramatically lowered the resistance and induced localized insulator-to-metal transitions. Remarkably, these pressure-triggered states proved to be the same ones as the much-discussed voltage-triggered memristive states in these materials [17].

Our motivating picture depicted in **Fig. 1(a)** is adapted from a simpler diagram of Street and Mott, who explained the energy variation of a negative-$U$ site in terms of its configurational coordinate [3]. The key idea is: While an extra electron usually raises the energy from A to B due to on-site Coulomb repulsion (positive $U$), by polarizing the neighboring environment so severely the electron may instead render a coordinate relaxation (bond contraction/extension) that lowers the overall energy to C. Since the relaxation realizes an effectively negative $U$, the extra electron becomes self-trapped [2]. As mentioned above, we hypothesized that such self-trapped electrons are blocking the few conducting paths in amorphous insulators. So, clearing them ought to render the insulator conductive. Therefore, our experimental objective is to find a way to block and unblock the conducting paths by leveraging the negative $U$ centers.

Our first experiment seeks to exert a mechanical force to restore the configurational coordinate from C to B, thus destabilizes the state and prompts spontaneous electron detrapping, back to state A, thereby clearing the conducting paths. In Anderson's example, this is like compressing an antibonding orbital to make it so unstable as to release the antibonded electron. In parallel, we will also exert an electrical voltage: (i) to raise the energy from C to D without disturbing their configurational coordinates, thus prompts electron detrapping after which the one-electron state relaxes back to A; and (ii) to raise the energy of A and B while keeping their configurational coordinates the same, thus adds more driving force for B to relax to C if B already captured a second electron. Indeed, (ii) may even work for a positive $U$, *i.e.*, when the energy of A is lower than that of C, although this will result in a metastable state C. All these pressure/voltage-triggered changes should leave very prominent signatures in resistance, as depicted in **Fig. 1(b)** by arrows that indicate transitions in the resistance spectrum: According to **Fig. 1(a)**, they are one-way transitions for pressure, and two-way ones for voltage. Here, we identify the resistance extrema as pristine insulator (P) where strong localization [18] reigns and metal (M) where weak localization [19] manifests, but we reckon that there may be other states of intermediate insulator (II) and intermediate metal (IM).

To lend support to the picture in **Fig. 1(a),** we have experimentally demonstrated all the transitions in **Fig. 1(b)**. Our experiments used $HfO_2$ and $Al_2O_3$ because they are strong and insulating materials, with band gaps from 5.7 to 8.7 eV [20], elastic moduli from 150 to 300 GPa [21,22] and melting points from 2,300 to 3,000K. In both bulk and nanoscale forms they are among the most reliable electrical insulators; amorphous $HfO_2$ films (1-2 nm) is the gate oxide in the state-of-the-art nanoelectronics capable of withstanding fields of ~ 1 V/nm [20,23]. For such application, amorphous films are especially advantageous because they are free of field-concentrating, breakdown-initiating lattice defects. The electric forces exerted by the breakdown field on the cations/anions are equivalent to a mechanical stress; in $HfO_2$ it is ~ 5 GPa. So a stress less than 0.5 GPa is unlikely to cause any significant defect formation, dielectric breakdown or phase transition. To further eliminate any artifact, we used a hydraulic pressure of no higher than 350 MPa. Indeed, as described below, we observed abrupt, dramatic and robust insulator-to-metal transitions in few-nm-thick amorphous $HfO_2$ and $Al_2O_3$ films even at 2 MPa.

Atomic-layer-deposition (ALD) is a method commercially used to fabricate conformal thin film gate oxides over a large area. We used it to deposit the above materials (5-15 nm thick), embedded in a metal-insulator-metal (MIM) structure between sputtered Pt film (40 nm) as top electrode and Pt or Ti as bottom electrode. The structure allows us to measure DC and AC current ($I$)-voltage ($V$) responses across the MIM to obtain the $V/I$ ratio as an indicator of resistance. The deposited Pt films have a grain size of 2 nm. Therefore, their finely spaced grain boundaries make it extremely difficult to nucleate and propagate dislocations. So they should have a tensile/compressive yield stress ~1 GPa according to the Hall-Petch relation [24,25]. This was verified in broken films: Unlike coarse-grain Pt that fails by grain tearing and thinning indicative of dislocation-mediated plasticity, our films broke in a completely brittle manner absent of any dislocation plasticity. This finding ensures that a pressure < 350 MPa cannot possibly cause any accidental shorting of the MIM by filling the oxide pinholes with Pt, because the pressure required to deform Pt to do so (like in an indentation test) is about three times the tensile/compression yield stress [26], thus exceeding 2 GPa. Below we will describe the Pt/$HfO_2$/Ti (thickness of $HfO_2$ being 10 nm) results in details; very similar findings for Pt/$HfO_2$/Pt and Pt/$Al_2O_3$/Ti MIM are presented in the Supplementary Material, **Table S1**.

The x-ray amorphous virgin films (data not shown) were all extremely insulating. For example, at ambient temperature a 172 μm radius Pt/HfO$_2$/Ti MIM has a resistance >100 GΩ read at 0.1V (left inset, **Fig. 2(a)**). As temperature is lowered, the resistance soon reached instrument's sensing limit (0.1 pA) as indicated by the dashed line. So a higher reading voltage was used to capture more features of the pristine state: The resistance rises with decreasing temperature $T$ (**Fig. 2(a)**) having the behavior of variable-range-hopping (right inset), but below ~160K elastic tunneling across the MIM caps the low-temperature resistance. For such as fabricated film—the P state in **Fig. 1(b)**—the Ohmic range is rather small before the resistance rapidly decreases with voltage, but the film does not suffer breakdown at least up to ±6 V when the *I-V* and *R-V* curves remain fully reversible.

A pressure treatment at ambient temperature was provided to pristine MIM arrays vacuum-sealed in an elastomer bag and suspended in a liquid-filled pressure vessel, which was charged and held at the set pressure for <5 min before sample removal (see inset of **Fig. 2(b)**). As shown in **Fig. 2(b)**, the brief treatment triggered a $10^8\times$ drop in resistance (from $10^{10}$ Ω to $10^2$ Ω) in this MIM, which acquired a totally different *R-T* dependence in 100-300K indicating a metal state (M). A shallow resistance minimum at $T_{min}$ ~40K also emerged at low temperature, which is a common feature of metal/bad-metal with impurities/disorders [27, 28] and a signature of weak (electron) localization [29]. (Hf$^{4+}$ has no magnetic moment, and our sample contains too few magnetic impurities, if any, to cause the Kondo effect.) The pressure-induced insulator-to-metal transition conforms to our expectation in **Fig. 1**: Pressure causes the C-to-B configuration/energy change, followed by the spontaneous B-to-A conversion, thus dumping the trapped electron, removing the Coulomb barriers and clearing the conduction paths. We also verified that the pressure treatment resulted in a continuum of intermediate states with different resistance values corresponding to the multiple blue arrows emanating from P in **Fig. 1(b)**. These states have distinctly different *R-T* curves in **Fig. 2(c)**, in which resistance *R* is normalized by their room temperature value $R_{300K}$. Here we distinguish II and IM states by the magnitude of $R/R_{300K}$ ratio: M has a ratio less than 10, which is the criteria commonly used to assign a bad metal instead of an insulator. In addition, as shown in Supplementary Materials **Fig. S1**, while II follows the $T^{-1/4}$ law of Mott [30] for variable range hopping over a relatively large resistance range before complete saturation in the tunneling regime at low temperature, the resistance of IM continues a gradual rise at the lowest temperature and the magnitude of change is just too small to invoke the notion of hopping.

The P – II – IM – M continuum resulting from the pressure-triggered transition of the pristine P state suggests a statistical phenomenon. This is evident from **Fig. 3**, which displays the Weibull plots of cumulative probability of the post-treatment resistance or the resistance ratio comparing pre- and post-treatment. (These probability plots having a sigmoidal shape correspond to the somewhat bimodal resistance distributions in Supplementary Material **Fig. S2.**) It clearly shows a wide distribution of the ratio, from 1 (a remaining P state) to $10^{-8}$ (an M state). As highlighted in **Fig. 3(a)**, pressure as low as 2 MPa, which is the accuracy of our pressure reading, can already trigger transitions in some MIM! But a higher pressure does give a higher transition yield; e.g., the cumulative probability, for ratio<$10^{-5}$, increases with pressure (**Fig. 3(a)**). Likewise, the cumulative probability increases in larger MIMs (**Fig. 3(b)**), again indicative of the statistical nature. Similar transitions probabilities were found in other MIMs shown in **Fig. 3(c)**, with a higher yield in $HfO_2$ than in $Al_2O_3$ perhaps because $HfO_2$ is not as stiff as $Al_2O_3$.

While the above experiments all started with the pristine P state, **Fig. 3(a)** also suggests a higher pressure can cause a II state to transition to another intermediate IM state, or even an M state. This was verified by pressure-re-treating a sample that was previously pressure-treated at a lower pressure (data not shown, but similar data will be shown later for magnetic pressure experiments.) We also verified that pressure cannot induce a metal-to-insulator transition: No M state can be pressurized into an IM, II, or P state, neither can an IM state or II state be forced by pressure to become more resistive. So all the blue arrows in **Fig. 1(b)** are indeed unidirectional. The irreversibility is not due to shorting, since we will describe later how a voltage, at room temperature or 2K, can cause the M-to-II transition in these pressured-transitioned M states. This is consistent with **Fig. 1(a)**, since it is the second electron that enters the electron-phonon interaction: Absent a self-trapped electron there is no interaction for pressure to tune.

One interesting feature of **Fig. 3(b)** is that their lowest resistances fall within a factor of 3-4 from each other despite their area varies as much as 25 times. This is inconsistent with a typical metallic state of a bulk, uniform nature. One possibility is that the pristine P state already has a filamentary conductive network that is nevertheless blocked at several critical points, and as the pressure treatment clears a few such points, the network finally becomes sufficiently conducting with its resistance mostly reflecting the local resistances around the few cleared-and-critical points in a way that is akin to resistance percolation [31-33]. (There are also those MIM that do not have enough blocked critical points cleared yet, so they remain in the P state, thus do not enter the statistics considered here.) In a rough estimation, we let the probability of pressure-clearing $n$ critical points to produce a conductive network to bridge the two electrodes in an MIM of area $A$ be proportional to $Aq^n$, where $q$ is the (pressure-dependent) probability to clear a single critical point. Therefore, for a given probability (i.e., at a given yield) and pressure, $n$ should vary with area as ln$A$ and the network resistance being reciprocally proportional to $n$ should follow 1/ln$A$, in agreement with the inset of **Fig. 3(b)**. (We envision $n$ parallel conducting paths cleared by $n$ points. These paths are connected with each other in the pristine network, but the connections may or may not be sufficiently conductive.) To seek further confirmation of the argument from an independent experiment, we reason that an increased oxide thickness, unlike an increased oxide area, should decrease rather than increase the yield because of the decreased probability for percolation in a longer network. Indeed, the average relative yield to reach $R$ ratio<$10^{-6}$ at 200 MPa in five sizes of Pt/HfO$_2$/Ti MIM in Supplementary Material **Fig. S3** was 30% for 5 nm thickness, 25% for 10 nm, and 17% for 15 nm. Lastly, if the conducting paths are along a filamentary network, then they are unlikely to have a large cross section. Indeed, despite hugely different resistance values, all the states in **Fig. 1(b)** do share a rather similar value of relative dielectric constant ~28 (see data in Supplementary Material **Fig. S4**), which is the same value seen in pristine HfO$_2$ [20]. These results are supportive of the picture of a percolative network. Further evidence of such network has been previously obtained in fracture experiments [34], which among other things can also afford a "peek" at the remnant sub-percolative network.

We already mentioned that pressure transitions are one-way transitions; pressure cannot induce any metal-to-insulator transition. But voltage can, according to **Fig. 1(a)**. To verify this and to address the issue of irreversibility of pressure transition, we conducted experiments to confirm all the voltage-enabled transitions depicted in **Fig. 1(b).** For brevity, we summarize the results of these experiments in **Table S1** and **Scheme S1**, and use **Fig. 4** to illustrate one such experiment. Here, we used a positive voltage of about 1 V (in our convention, a positive voltage forces current to flow from the top electrode to the bottom electrode) to convert M—obtained from a previous pressure treatment—to II, then used a negative voltage of about −1 V to convert II to M. We also used a voltage to convert P to M, and next, M-to-II. In this way, we verified all the green arrows in **Fig. 1(b)**. From these results, we conclude that the irreversibility of pressure transition is intrinsic as expected from **Fig. 1**, and not because the MIM was damaged by the pressure treatment.

Note that, once again, we cannot convert M, IM, or II into P. So, once the P state has transitioned to a lower resistance state, it cannot be recovered. This is not unreasonable because statistically it is very unlikely to reestablish (by electron trapping) same blocking at all the critical points—any failure to do so at just one critical point is likely to lower the network resistance by a large amount, hence the impossibility of returning to the P state.

We next compare the two M states obtained by pressure and voltage treatment, respectively. First, we verified their resistances are similar and insensitive to their respective cell area, implicating the same type of metallic network states in both. Next, we compared their subsequent voltage-induced two-way transitions and found them essentially indistinguishable (**Fig. 4**), not only at 300K but also at 2K. Note in particular that the transition voltage is insensitive to temperature, which is consistent with the picture in **Fig. 1(a)** where the transition criterion is based on energy only. Lastly, when the two II states were subject to a pressure treatment, they exhibited the same transition statistics in **Fig. 3(d)** and Supplementary Material **Fig. S5**, again leaving a continuum of intermediate states and again with a pressure-dependent transition yield (**Fig. S5**). However, their transition statistics are insensitive to the cell area unlike the case in **Fig. 3(b)**. This is a general observation: The area dependence only appears in the first transition from the P state and not in the subsequent pressure-induced transitions. This can be understood in the context of the network picture as follows. A comparable resistance of M or II implies the same (or comparable) number of unblocked critical points. Any subsequent transition is thus mostly likely to involve the same set of critical points; a voltage-induced M-to-II transition re-blocks some of these points, and a voltage- or pressure-induced II-to-M transition unblocks some of these points. Inasmuch there is no longer a need to involve new points—the random statistics of finding them is the origin of the area dependence—there is no longer an area dependence. In this way, all the transitions depicted in **Fig. 1(a)** have been verified, and all the data displayed in **Fig. 3-4** including their statistics can be rationalized.

This rich variety of transitions is summarized in **Scheme S1** and **Table S1**, where we summarize all the transitions. There is clearly an analogy between pressure and voltage, except that pressure transitions are one-way whereas voltage transitions are two-way. This also applies to other MIMs having a different electrode configuration (Pt/Pt *vs*. Pt/Ti) or a different oxide type ($Al_2O_3$ *vs*. $HfO_2$) as shown in **Table S1**. Since they all provide the same findings, electrode-induced redox reactions (Ti is a reducing agent but Pt is not) and bulk polarizability ($HfO_2$'s dielectric constant is 4× that of $Al_2O_3$) cannot be a major factor in these transitions. These transitions leave very similar states in **Scheme S1**, with very similar resistances as well as their temperature dependence. In all cases, the metallic and weak-localization features of the M state and the insulating feature of the II state are clearly manifest.

Previously, irreversible initial voltage transition from the pristine state and subsequent reversible voltage transition between two resistance states were observed in amorphous $HfO_2$, $Al_2O_3$, and many other oxides but interpreted entirely differently emphasizing oxygen ion migration that causes a "soft" dielectric breakdown along a filament and subsequent electrical reconnection of the filament, possibly involving local Joule heating [17,35,36]. Such consensus mechanism obviously cannot explain pressure-induced insulator-to-metal transitions in which neither voltage nor heat was involved. On the other hand, crystalline $HfO_2$ and $Al_2O_3$ are likely to be Mott insulators [37] with strong electron correlation that forbids electron delocalization, which rules out the metallic state unless there is a global structural change or electron/hole doping. So it is very interesting that their disordered oxides at such modest pressure can undergo an insulator-to-metal transition. The transition is more reminiscent of a crossover from strong localization to weak localization involving a few local changes of local random potential or bandwidth [38], albeit in $HfO_2$ and $Al_2O_3$ thin films the conducting paths are apparently network-restricted and not widespread. As mentioned in the introduction, we believe these local changes occur at the soft spots in the disordered structure, and their strong local deformability, hence large polarizability, stabilizes trapped electrons by electron-phonon interaction. The few conducting paths that constitute the network and are blocked by these trapped electrons may simply reflect the local statistical variations in bandwidth and random potential of an amorphous structure in which the possibility of finding a few short-distance conducting paths cannot be excluded.

Lastly, we describe a subpicosecond experiment designed to give further evidence for the electron-phonon mechanism. Physically, the conversion from configuration C to configuration B in **Fig. (1a)** may take as short as the period of one atomic vibration, ~0.1 ps, which sets the upper limit for the speed of the insulator-to-metal transition. We have previously reported how an ultrafast, sub-picosecond pressure impulse can be provided by a Lorentz switch that harnesses the magnetic transient of an ultrarelativistic electron bunch to mechanically induce an insulator to metal transition by locally reversing a negative $U$ [39]. We have verified that this switch also works for $Pt/HfO_2/Ti$ MIM giving the same results as reported for $Pt/Si_3N_4:Pt/Mo$ except the transition here is network-percolative. (In $Pt/Si_3N_4:Pt/Mo$, it is an apparently uniform, bulk like transition.) Briefly, the experiment was performed inside a linear accelerator (SLAC) that shot, just once, a single bunch of $10^9$ 20 GeV electrons at the film. The bunch has a size of 20 μm×20 μm×20 μm and carries with it a $10^{-13}$ s pulse of circumferential magnetic field, the pulse time being the time to travel 20 μm at the speed of light [40]. The field peaks at 65 T at the edge of the bunch ($r=20$ μm), falls off as $\sim 1/r$ with the radial distance $r$, and remotely delivers to a two-side-electroded film (located at $r$ away from the flight path) a magnetic pressure $P_B$ that peaks at 1,680 MPa and decays roughly with $\sim 1/r^2$. (There is also an induced pressure due to the induced current in the electrodes, which inversely depends on the (pulse width)$^2$)[38]. As explained in the schematic of **Fig. 5(a)** and the caption, the pressure generates a uniaxial tension to stretch the oxide film. It also generates a biaxial tension in the top electrode, which can be ripped apart if the electrode is thin, but not if the electrode is thick. (The stiffness increase with the third power of the thickness.) This is evident in Supplementary Material **Fig. S6** in which the damage zone caused by the electron bunch shrinks as the electrode thickness increases. This feature makes it easy to identify where the flight path hit the MIM array, and since the electron bunch is only as big as an MIM cell, ~20 by 20 μm$^2$, each of them appearing as a "dot" in **Fig. 5(b)** and Supplementary Material **Fig. S7**, all the pressure impulses seen by the cell are entirely remote and magnetic in origin, and they radially decay from the center (where electron bunch hit) that is indicated by a radial arrow. Per reversible transition II-M curve in **Fig. 5(c)**, we often also preset some MIM cells to II and M states prior to the experiment to study various transition possibilities. In the following, we describe three key observations of these 0.1 ps pressure-impulse-transition experiments.

(a) One-way transitions: In an MIM array that contains either P states (red) or voltage-preset II (yellow) and M states (medium blue) shown on the left of **Fig. 5(b)**, we found after one single shot of an electron bunch, resistances of cells within an area from the flight path were lowered to mostly that of M states, shown on the right of **Fig. 5(c)** which contains numerous blue dots. Next we examine the three exemplary initially P, II and M states, marked by the white circles and located on a circle of $r$~300 μm from the center, thus having seen the same magnetic pressure. While the M cell remained as M but the resistance decreased from 200 Ω to 100 Ω, both P and II cells had transitioned to the M state. This confirms that the magnetic pressure is effective to induce all the one-way transitions depicted in **Fig. 1(b)** despite its subpicosecond duration. The critical pressure can be estimated by the edge of the "blue circle", which extends to ~300 μm, where the primary magnetic pressure is estimated to be 10 MPa.

(b) Percolative transition: Supplementary Material **Fig. S7** shows another MIM array that has thinner top electrodes, which were peeled off by the magnetic pressure when the cells are located too close to the center. A few cells, however, still had some remnants of the top electrodes, and two such cells are marked by white circles. Although they have about the same radial distance from the center, hence the same magnetic pressure, they have very different resistances (2 GΩ vs. 300 Ω). Comparable tests that used MIM arrays of thicker electrodes found that, at this radial distance, all the cells should have transitioned to the M state, so the highly resistive cell in this set seems to be an anomaly. However, such anomaly is consistent with the picture of percolative network because if a cell had lost the part of the top electrode that was in contact with the percolative part of the network, then the remaining part of the top electrode would not be able to communicate with the bottom electrode, hence such cell will read a very high resistance. A check of other cells with similar top-electrode remnants confirmed this picture: Their resistances are either that of the M state, or very much higher. For the 10 cells at $r$=250 μm with electrode remnants in Supplementary Material **Fig. S7**, 4 of them have resistance ~ 300 Ω, 2 cells have resistance ~ 10 MΩ, and the other 4 cells have resistance ~ 2 GΩ. This finding is similar to a previous report of ours, in which we intentionally severed the cells to study the resistance statistics of the two severed halves [34].

(c) Cell reversibility: To again check whether the magnetic-pressure-transitioned M states were damaged or not, we applied an electrical voltage to see if they could be transitioned to II, and indeed they could, with a voltage transition curves that appear normal (red curve in **Fig. 5(c)**). This not only rules out any permanent damage caused by the magnetic pressure but also once again establishes the equivalency between the pressure-transitioned state and the voltage-transitioned state. Thus, there is no difference between the two pressure experiments, one by a static hydraulic pressure and the other a dynamic 0.1 ps magnetic pressure: They both reverse the sign of negative $U$ in the same way.

In summary, we have discovered a pressure-induced nanoscale insulator-to-metal transition in nanofilms of amorphous $HfO_2$ and $Al_2O_3$ in both pristine form and voltage-conditioned form, and such transitions are proffered as evidence of negative-$U$ centers in these materials. The discovery is significant not only because the transitions are extremely dramatic involving a $10^9$-fold drop of resistance and requiring just a $10^{-13}$ s pressure impulse, but also because they occur in such wide band-gap and low polarizability materials that have not been thought of as natural hosts of negative-$U$ centers and strong electron-phonon interactions. On another front, our results strongly suggest that Mott insulators, once disordered, become Anderson insulators, and within the latter, there is a possible crossover from strong localization to weak localization as the sample size is reduced to the nanoscale where a short conducting length of the order of 10 nm becomes important. Possibly, the resultant incipient conduction paths capable of metal-insulator transitions are one-dimensional Fermi glasses worthy of further studies. Obviously, these findings also carry considerable technological potential.

**Methods**

*Sample Preparation*: MIM structures of amorphous oxides were fabricated on a substrate of thermal-oxide-coated 100 oriented *p*-type silicon single crystals. For bottom electrode, 15 nm Ti bottom electrode was first deposited by e-beam evaporation or DC sputtering and 20 nm Pt was next deposited by RF sputtering. After that, 5/10/15 nm $HfO_x$ or $AlO_x$ layers were deposited by ALD at 250°C using $H_2O$ with tetrakis(dimethylamido)hafnium (HFDMA) or Trimethylaluminum (TMA) precursor. Finally, a 40 nm thick Pt top electrode was deposited by RF sputtering either through a shadow mask that defined cells of 50-250 μm in radius, or onto a lithography-defined pattern of 1-20 μm radius cells followed by a lift-off process. The former type of cells was used for hydraulic pressure experiments and transport measurements, while the latter type was used for most magnetic pressure experiments.

*Electrical Measurement*: Electrical properties were measured on a Signatone S-1160 probe station. In a typical test configuration, a bias voltage was applied to the top Pt electrode while the bottom contact was grounded. DC current–voltage (*I-V*) characteristics were examined using a semiconductor parameter analyzer (SPA, Keithley 237). AC impedance measurements were performed using a HP4192A impedance analyzer ($10^2$-$10^7$ Hz).

*Hydraulic Pressure Experiment*: Before pressure treatment, resistance of each pristine cell in MIM arrays was either read at 0.2 V or pre-switched to certain resistance state after checking for voltage-induced transitions using Keithley 237. Then they were covered by aluminum foils, vacuum-sealed in elastomer bags, and suspended in a liquid-filled isostatic pressure vessel (Autoclave Engineers, Erie, US), which was charged to a pressure at room temperature and held for <5 min before sample removal. After the pressure treatment, the resistance of each cell was read again at 0.2 V and its voltage-induced transition curve recorded. The reading and curve were compared with their pre-treatment counterparts.

*Low Temperature Experiment*: Electrical data were collected in a cryostat (PPMS of Quantum Design, San Diego, US.) Samples were mounted on a special chip holder with a heat conducting vacuum grease. Silver paint was used to bond gold wires (dia.=2 μm) to device electrodes and to connect to the pins on the sample holder. During temperature sweeping from 300 K to 2 K, resistance of MIM cells in different states of II, IM and M was read at 0.01 V (P state read at 0.1 V and 1V due to the instrumental limit) using Keithley 237. *I-V* or *R-V* transition curves were also obtained at certain temperature by DC voltage-sweeping.

*0.1 ps Impulse Magnetic Pressure Experiment:* A magnetic pressure burst was used to trigger insulator-to-metal transition. In principle, a magnetic pressure may be generated by passing a burst magnetic flux through an insulator-filled gap between two electrodes, which form a metallic "container" that confines the burst magnetic field. (At high frequency, the gap resistance at the edges of the electrodes is very small so the two electrodes form a continuous circuit.) Assuming $HfO_2$ has a relative permeability of unity, the magnetic pressure $P_B$ is $(B/0.501)^2$ with the pressure expressed in bar (1 bar=0.1 MPa) and *B* in T. The burst magnetic flux was received from an electron bunch, which is a spatially localized bundle of 20 GeV electrons generated at Stanford Linear Accelerator Center (SLAC) using the FACET facility. Each bunch contained ~$10^9$ electrons (>1 nC) that are narrowly collimated (~20 μm). It had a short duration, passing in ~0.1 ps, which is the time for the bunch to travel 20 μm at near the speed of light, and it was available on a bunch-by-bunch basis. We only allowed each cell to see one bunch during the experiment; after each shot the MIM array was moved to a new position sufficiently away from the first location before a second shot was fired. The electron bunch hit the array chip in the normal direction. Since the maximum magnetic field around the bunch is ~ 65 T at the edge of the bunch, i.e., ~ 20 μm from the flight path, and it decays with the radial distance $r$ from the bunch roughly according to $1/r$, we can estimate the magnetic pressure in each cell from the cell location relative to the flight path. (To maximize the induced magnetic pressure inside the cell, we chose the cell size to be 20 μm, comparable to the bunch size.) The estimated magnetic pressure at 250 μm away is ~20 MPa, which is a lower bounds since it does not consider the pressure caused by the induced current in the electrodes. The electric field is radial and follows the same radial variation as the magnetic field, but it is unimportant for this experiment as previously established in Ref. 39.

Before the magnetic-pressure treatment, cells were either left in their pristine state or pre-transitioned by a voltage to certain resistance states, with their two-point resistance values recorded at 0.1 V by a Keithley 237. Their resistance was again read in the same way after the magnetic-pressure treatment and compared with the pre-treatment value. In a typical representation of the data, each cell is colored to indicate its resistance value before and after the treatment, and the colored maps are presented to aid comparison. Since the cell size is about the same as the bunch size, maps that have hundreds of cells—each appearing as a "dot" with changed colors due to the treatment—provide direct evidence for the far-field effect of an electron bunch.

**Acknowledgements**

This research was supported by the US National Science Foundation Grant No. DMR-1409114. The use of facilities including PPMS at Penn's LRSM supported by DMR-1120901 is gratefully acknowledged. Work was also performed at SLAC National Laboratory supported by the US Department of Energy, Office of Basic Energy Sciences. Experimental assistance of Prof. Jay Kikkawa (Penn) and Ioan Tudosa (SLAC) is gratefully acknowledged.

18. P. W. Anderson, *Phys. Rev.* **109,** 1492 (1958).

19. D. M. Basko, I. L. Aleiner, and B. L. Altshuler, *Ann. Phys.* **321,** 1126-1205 (2006).

20. G. D. Wilk, R. M. Wallace, and J. M. Anthony, *J. Appl. Phys.* **89,** 5243-5275 (2001).

21. J. B. Wachtman Jr, W. E. Tefft, D. G. Lam Jr, C. S. Apstein, *Phys. Rev.* **122,** 1754 (1961).

22. R. Thielsch, A. Gatto, N. Kaiser, *Appl. Opt.* **41,** 3211-3217 (2002).

23. A. I. Kingon, J. P. Maria, S. K. Streiffer, *Nature* **406,** 1032-1038 (2000).

24. T. G. Nieh, and J. Wadsworth, *Scripta Metall. Mater.* **25,** 955-958 (1991).

25. N. Hansen, *Scripta Mater.* **51,** 801-806 (2004).

26. A. L. Romasco, L. H. Friedman, L. Fang, R. A. Meirom, T. E. Clark, R. G. Polcawich, J. S. Pulskamp, M. Dubey, & C. L. Muhlstein, *Thin Solid Films*, **518,** 3866-3874 (2010).

27. M. A. Howson, and B. L. Gallagher, *Phys. Rep.* **170,** 265-324 (1988).

28. P. Sheng, *Introduction to wave scattering, localization and mesoscopic phenomena* (Vol. 88). Springer Science & Business Media (2006).

29. P. A. Lee, T. V. Ramakrishnan, *Rev. Mod. Phys.* **57,** 287–337 (1985).

30. Mott, N. F., & Davis, E. A. *Electronic processes in non-crystalline materials*, 2nd ed, Oxford University Press, pp. 32-37 (1979).

31. V. Ambegaokar, B. I. Halperin, and J. S. Langer, *Phys. Rev. B*, **4,** 2612 (1971).

32. B. I. Shklovskii, and A. L. Efros, *Sov. Phys. JETP*, **33,** 468 (1971).

33. M. Pollak, *J. Non-Cryst. Solids*, **8,** 486-491 (1972).

34. Y. Lu, J. H. Lee, X. Yang, X., and I. W. Chen, *Nanoscale*, **8,** 18113-18120 (2016).

35. R. Waser, R. Dittmann, G. Staikov, and K. Szot, *Adv. Mater.* **21,** 2632-2663 (2009).

36. H. S. Wong, H. Y. Lee, S. Yu, Y. S. Chen, Y. Wu, Y., P. S. Chen, B. Lee, F. T. Chen and M. J. Tsai, *J. Proceedings of the IEEE* **100,** 1951-1970 (2012).

37. M. Imada, A. Fujimori, and Y. Tokura, *Rev. Mod. Phys.* **70,** 1039 (1998).

38. P. W. Anderson, *Phys. Rev.* **109,** 1492 (1958).

39. X. Yang, I. Tudosa, B. J. Choi, A. B. Chen, I. W. Chen, *Nano Lett.* **14,** 5058-5067 (2014).

40. S. J. Gamble, M. H. Burkhardt, A. Kashuba, R. Allenspach, S. S. Parkin, H. C. Siegmann, and J. Stöhr, *Phys. Rev. Lett.* **102,** 217201 (2009).

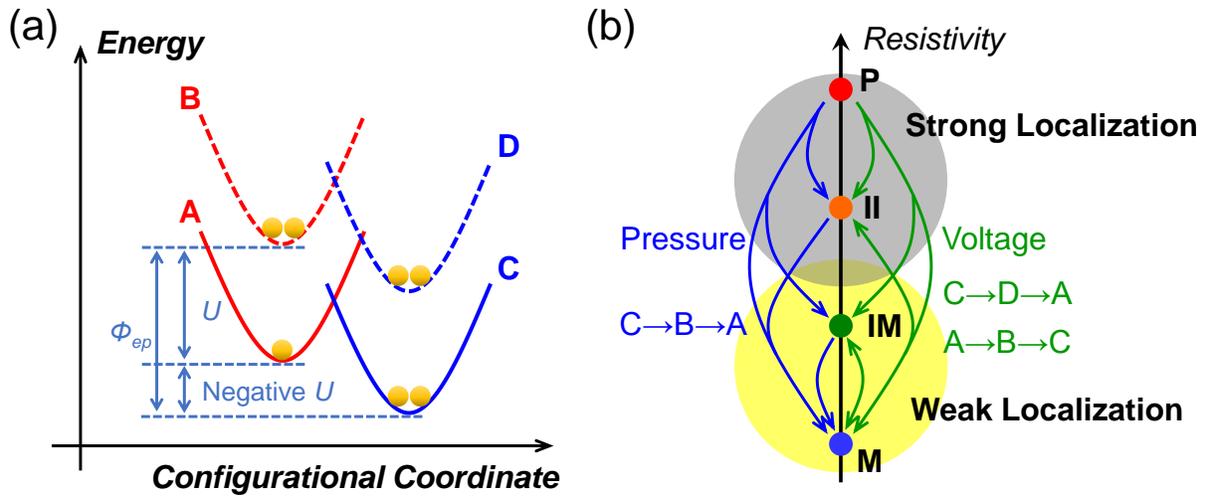

**Figure 1 Mediated by electron-phonon interaction $\phi_{ep}$, negative-$U$ state regulates charge-trapping/detrapping and enables pressure/voltage-induced metal-insulator transitions over a wide spectrum from strong to weak localization.** (**a**) A: Empty trap-state (single-electron occupancy, conducting); B: Filled trap-state (double-electron occupancy, insulating) without relaxing configurational-coordinate, showing positive $U$; C: same as B but with relaxed configurational-coordinate, showing negative $U$; D: same as C but voltage-elevated to metastable state without relaxing configurational-coordinate. Although only one A state and one C state are shown, in amorphous materials there are a multitude of A states and C states. (**b**) Pressure and voltage induced transitions between pristine insulating state (P) and metallic state (M), via a intermediate continuum including intermediate metallic state (IM) and intermediate insulating state (II). For pressure transitions (one-way, shown in blue), P-to-M/IM/II, II-to-M and IM-to-M transitions in (b) correspond to reaction C→B→A in (a). For voltage transitions (two-way, shown in green), P-to-M/IM/II, II-to-M, and IM-to-M transitions in (b) correspond to C→D→A in (a); and M/IM-to-IM/II transition in (b) corresponds to A→B→C in (a). All transitions have been experimentally verified in this work.

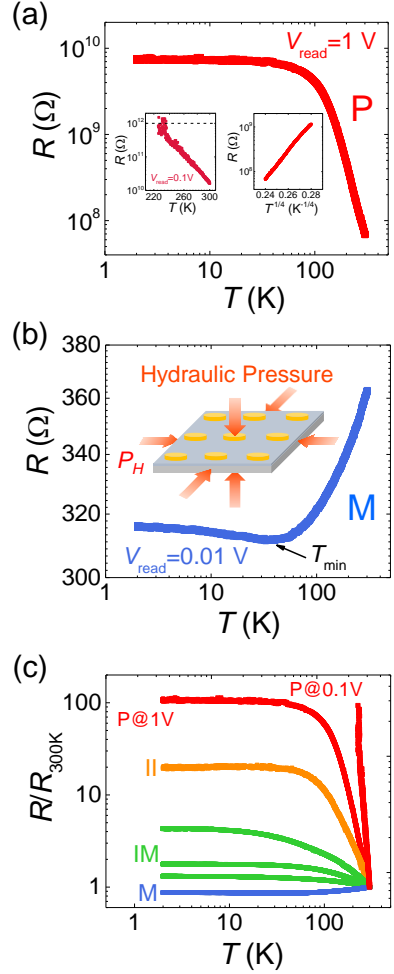

**Figure 2 Pressure induced insulator-to-metal transition in amorphous $HfO_2$.** **(a)** As-fabricated (pristine) Pt/$HfO_2$/Ti MIM resistance ($R$) read by voltage $V_{read}$ =1 V *vs*. temperature ($T$). Flat top is elastic tunneling limit. Left inset: Resistance read at 0.1 V curve is much higher, reaching instrument's limit. Right inset: variable range hopping plot, $\log R \sim T^{-1/4}$. **(b)** Post-pressurization (350 MPa) $R$-$T$ curve read at 0.01 V has low resistance, which is voltage-independent (Ohmic, not shown) with a shallow minimum at $T_{min} \sim 40$ K—a common feature of disordered electron systems. Inset: schematic of applying hydraulic pressure to a MIM array. **(c)** Normalized by their 300K resistance, various post-pressurization $R$-$T$ curves demonstrate intermediate states obtainable from pristine Pt/$HfO_2$/Ti MIM. (Cell size: 172 μm; pressure: 350 MPa). One M (a different one from **(b)**) with $R_{300K}$=162 Ω; three IM with $R_{300K}$=505 Ω, 672 Ω, and 865 Ω; one II with $R_{300K}$ = 25 kΩ, all the above read at 0.01 V; one P with $R_{300K}$ = 10 GΩ read at 0.1V and 68 MΩ read at 1V. IM and II distinguished by setting $R/R_{300K}$ (<10 for IM as commonly so designated for dirty metals) and $\log R \sim T^{-1/4}$ plot (applicable over wide resistance range for II) in **Fig. S1**.

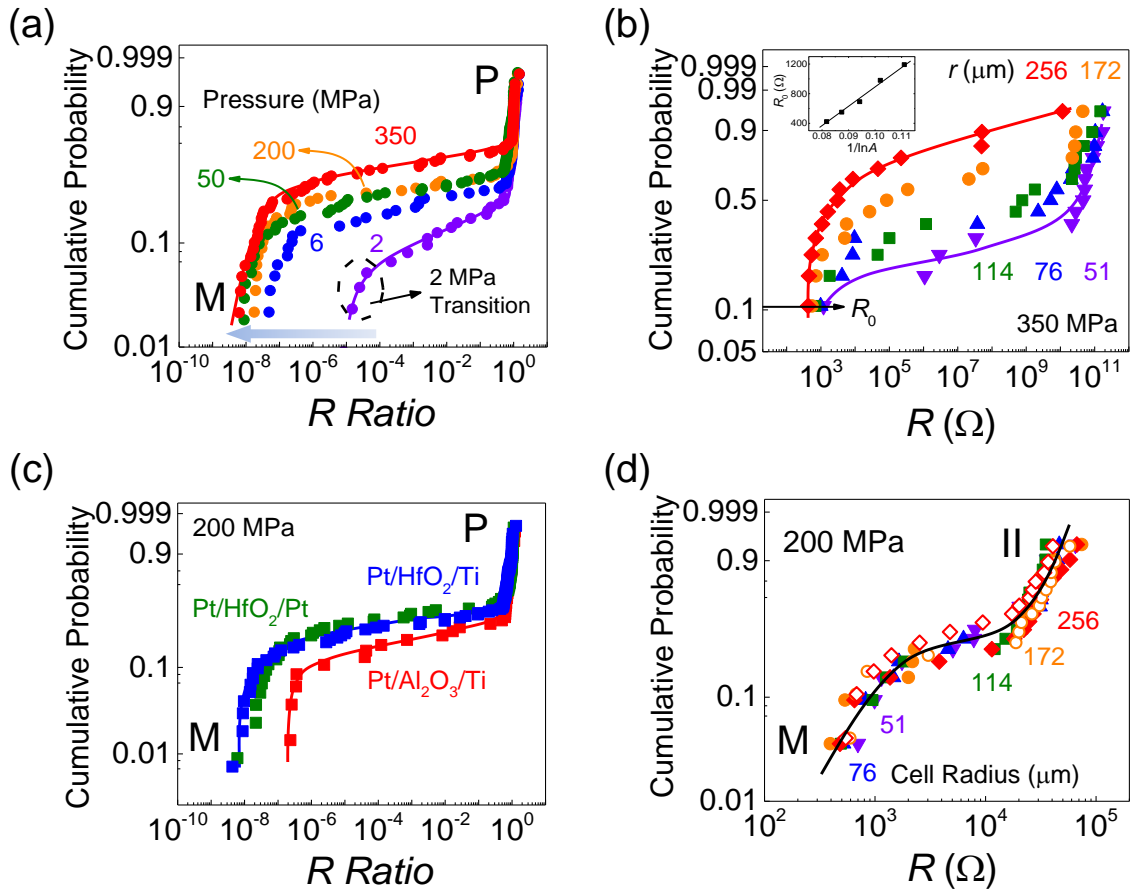

**Figure 3 Statistics of pressure-induced insulator-to-metal transition in HfO$_2$.** Data shown in Weibull plots of cumulative transition probability $p$ in $\ln[\ln[1/(1-p)]]$ scale *vs.* resistance ratio (post-pressurization to initial) or post-pressurization resistance, where resistance are read at 0.2 V and solid curves are guides to the eye. (**a**) Weibull plots showing higher $p$ at higher pressure. Each curve contains 60-80 pristine cells (10 nm thick), equally distributed between 5 cell sizes $A$ (with radius $r$). Sigmoidal plot signifies a bimodal resistance distribution (see **Fig. S2**) with two modes at two ends of P and M states. Dash-circled cells transitioned at 2 MPa. (**b**) Weibull plots of P-to-M transition showing higher $p$ at larger cell radius $r$ pressurized at 350 MPa. Inset: Post-pressurization resistance $R_0$ at $p$=10.5% (marked) is proportional to $1/\ln A$. (**c**) Weibull plots showing similar transition curves for different MIM configurations, all from pristine state under 200 MPa. Different electrode configurations for HfO$_2$ have the same $p$, which is higher than for Al$_2$O$_3$. (**d**) Weibull plots for pressure-induced *II*-to-*M* transition of pressure-initiated (hollow symbols) and voltage-initiated (solid symbols) cells: A pressure/voltage was used to initiate transition from P to M, then a voltage was used to transition M to states to II, followed by pressurization at 200 MPa. Both share similar $p$ and bimodal resistance distribution, and, unlike (b), no area dependence.

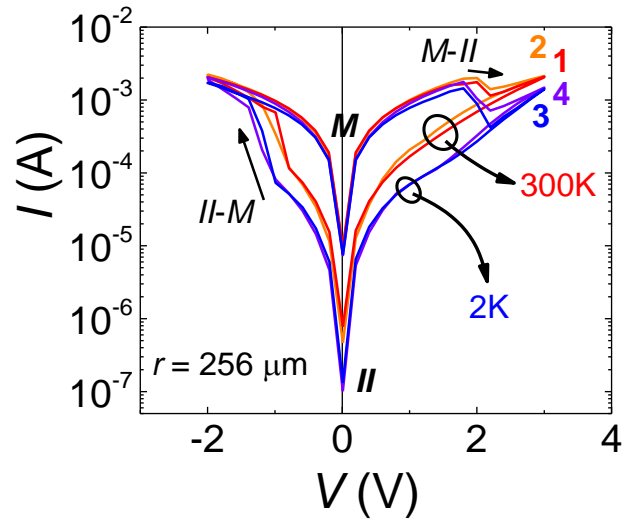

**Figure 4 Voltage-induced two-way transition between II state and M state in HfO$_2$.** Bipolar current-voltage (*I-V*) M-II-M transition loops of either pressure-initiated M state (loop 1, 3) or voltage-initiated M state (loop 2, 4), initiation starting with two pristine MIMs of same cell radius (256 μm). As marked by circles, resistance values of II differ in loops at 300K (loop 1, 2) and 2K (loop 3, 4), in contrast to similar resistance of M in all loops. Arrows indicate switching directions of M-to-II (+ bias) and II-to-M (− bias).

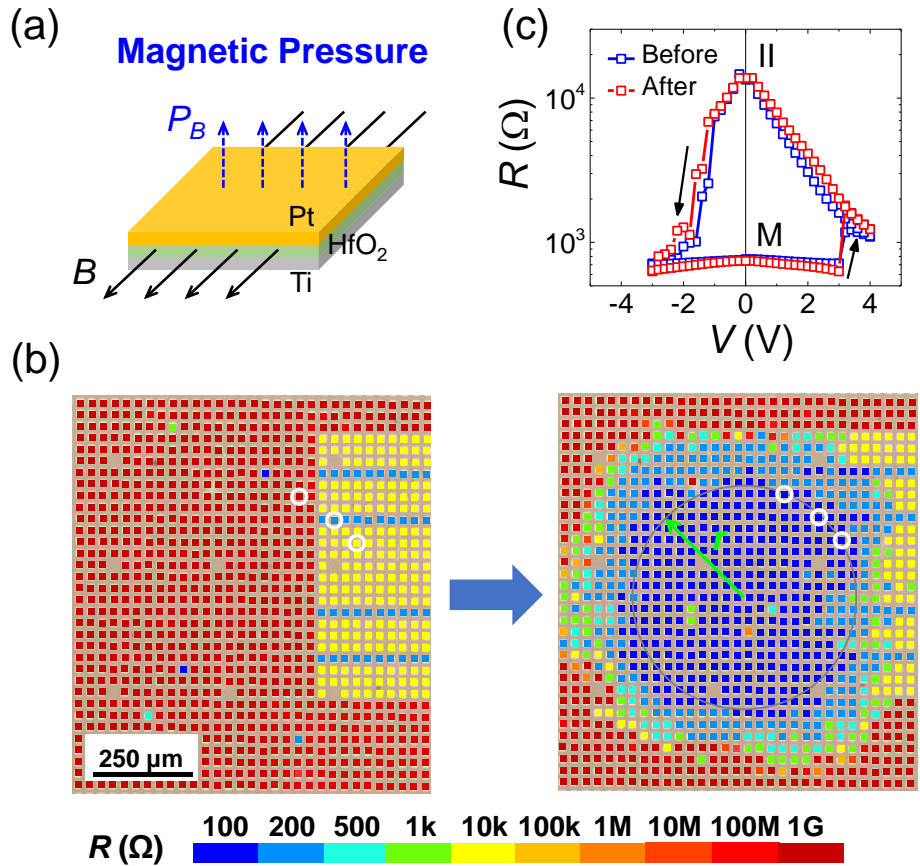

**Figure 5 Subpicosecond insulator-to-metal transitions in $HfO_2$, all of them simultaneously induced by one $10^{-13}$ s impulse of magnetic pressure.** (a) Schematic of the magnetic pressure $P_B$, exerted by remote magnetic field $B$ of electron bunch, which is a body force on the electrode conductors. To keep the upper electrode in balance an interface uniaxial tension of the same magnitude as $P_B$ must be generated, which will also stretch the insulating film like a negative pressure. (b) Left: Most $Ti/HfO_2/Pt$ MIM cells left in pristine P states (red), some voltage-preset to II (yellow) and M (medium blue), which are interchangeable by voltage transition following red transition R-V curve (marked as before) in (c). Each colored "dot" represents one cell. Right: After electron bunch (of a size of 20 μm) hit center ($r$=0) once, cell resistances are re-measured and cells are colored per color spectrum at bottom. As exemplified by the three cells marked by white circles at same distance ($r$=300 μm) from center, P and II are transitioned to M state, and M state of 200 Ω lowers resistance to 100 Ω. Later, these impulse-magnetic-pressure transitioned M states can be voltage-transitioned to II following blue transition R-V curve (marked after) in (c), which appears the same as red curve before.

# Supplementary Materials

**Table S1. Data summary of Pt/HfO$_2$/Ti, Pt/HfO$_2$/Pt and Pt/Al$_2$O$_3$/Ti MIM. (IM state omitted for brevity.)** The pristine state (P) can be pressure- or voltage-transitioned to metallic state (M), then voltage-transitioned to intermediate insulating state (II), which can again be pressure-transitioned to M state.

| MIM Structure | Pt/HfO$_2$/Ti | Pt/HfO$_2$/Pt | Pt/Al$_2$O$_3$/Ti |
|---|---|---|---|
| **Virgin Resistance** | 20 GΩ @0.2 V | 15 GΩ@0.2 V | 20 GΩ@0.2 V |
| **Pressure-induced P-to-M** | 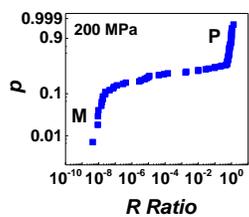 | 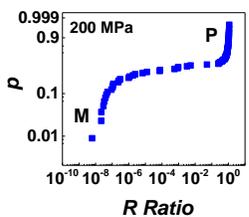 | 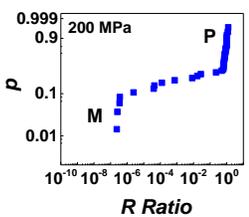 |
| **Voltage-induced P-to-M** | 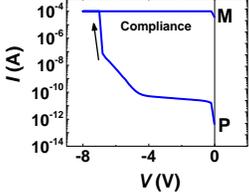 | 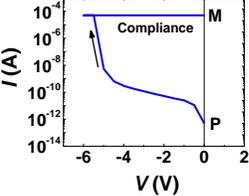 | 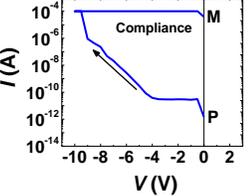 |
| **Voltage-induced II-to-M and M-to-II** | 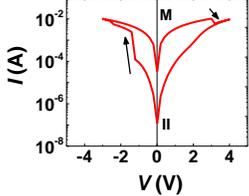 | 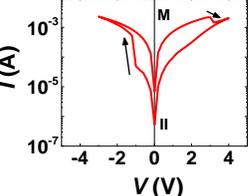 | 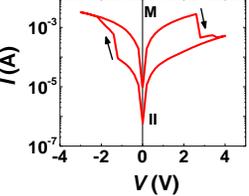 |
| **Pressure-induced II-to-M** | 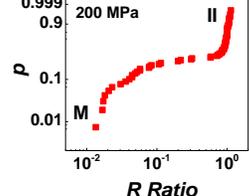 | 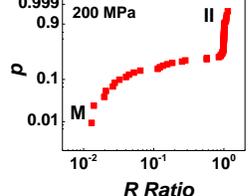 | 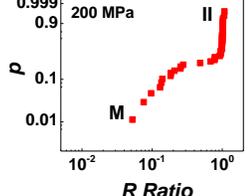 |

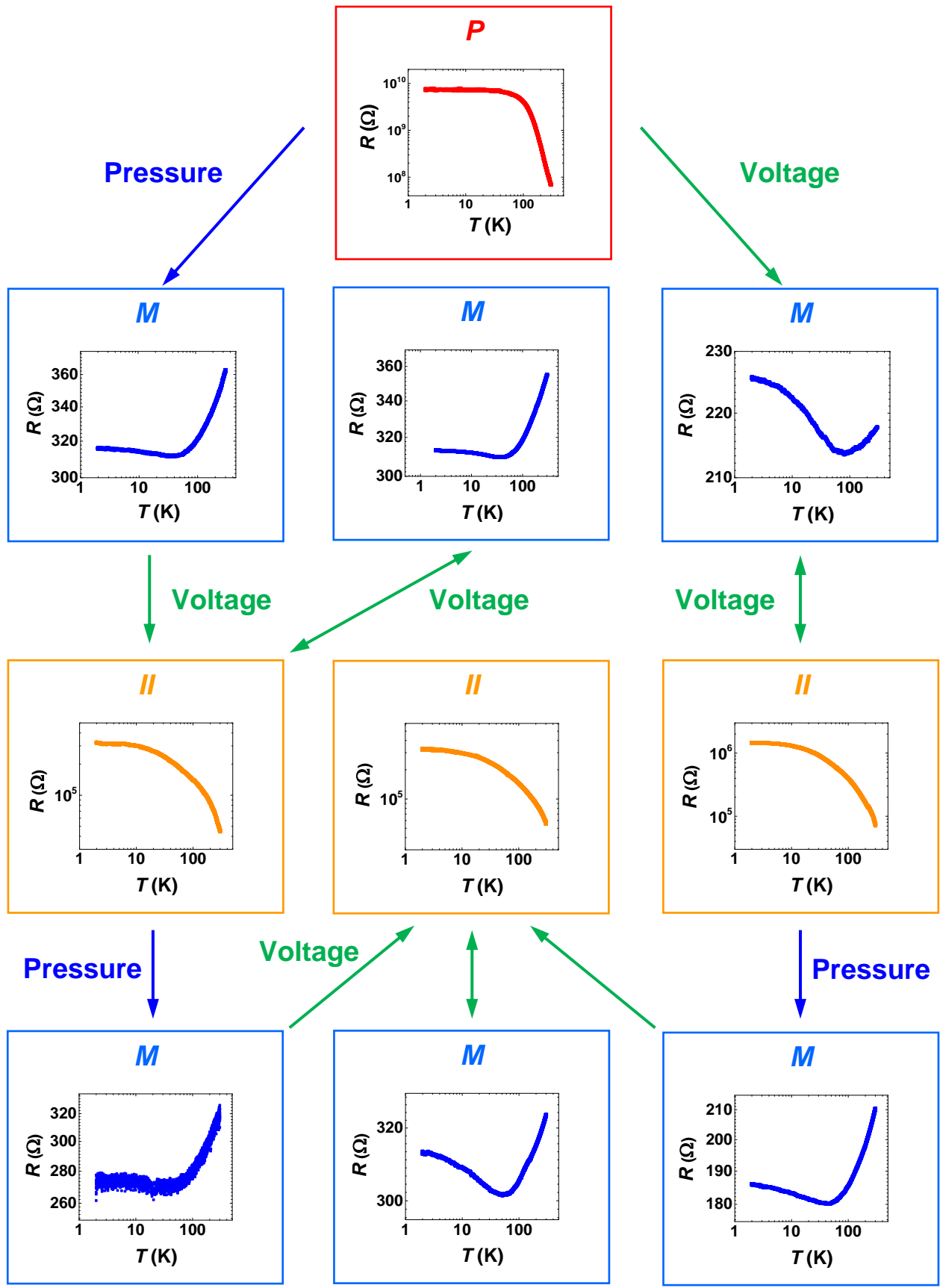

**Scheme S1. Summary of *R-T*-curves in different states, starting from pristine state through various possible pressure/voltage treatments. (For brevity, IM states are not included.)** Blue arrows indicate pressure-induced one-way insulator-to-metal transitions to metallic state M from either pristine state or intermediate insulating state II. Red arrows indicate voltage-induced one-way transition to M or two-way transitions between M and II. All P/II curves feature variable-range-hopping at high temperature and elastic tunneling at low temperature; all M curves feature resistance minimum at some $T_{min}$. Resistance values read at 0.01 V except for P, read at 1 V.

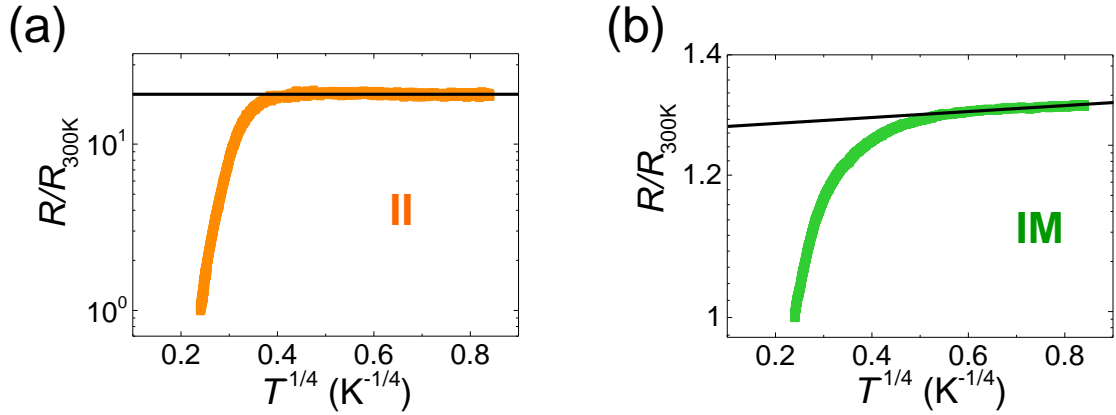

**Figure S1 R-T$^{-1/4}$ plots for intermediate insulator (II) and intermediate metal (IM) states of HfO$_2$**. **(a).** Resistance of II follows Mott's log$R \sim T^{1/4}$ relation at high temperature and saturates at elastic tunneling resistance at low temperature. **(b).** Resistance of IM varies within limited range insufficient to verify Mott's law even though it continues to rise slowly at low temperature.

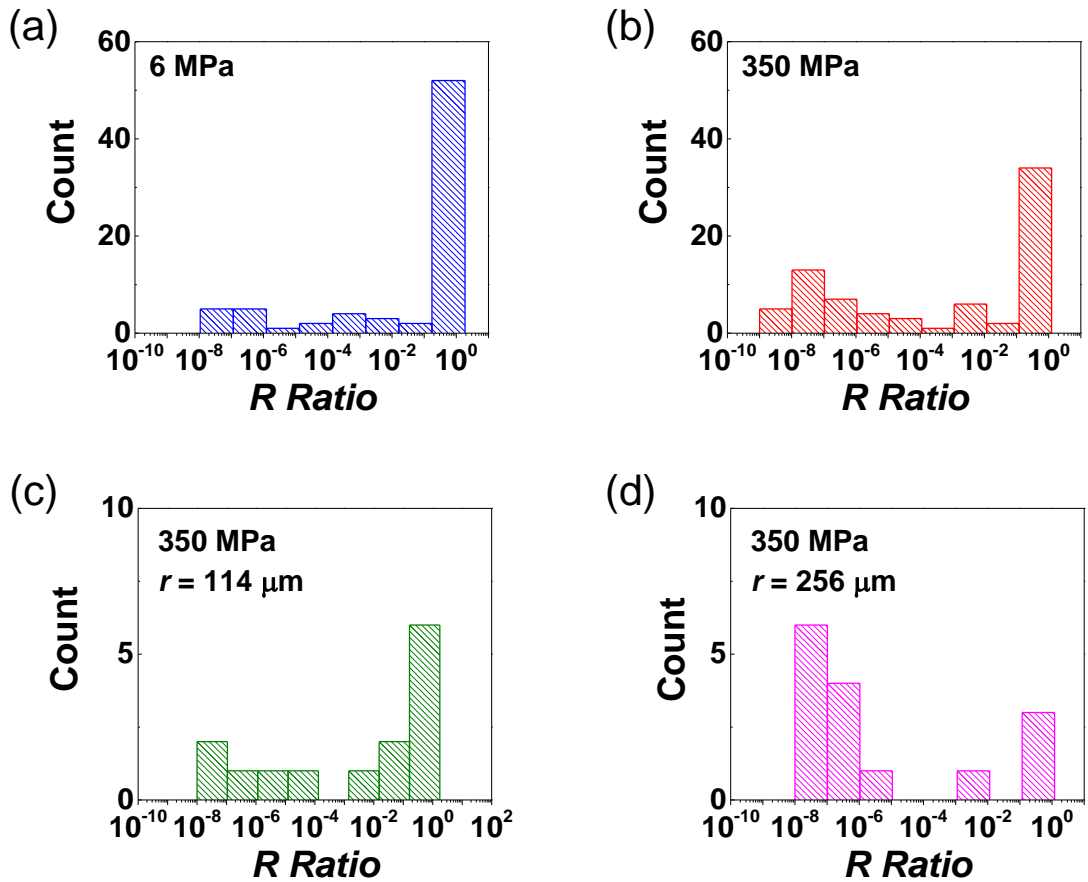

**Figure S2 Bimodal resistance distributions of pressure-transitioned states**. Histograms of $R$ ratio (post-pressurization to initial) for **(a)** 6 MPa data with equally distributed 5 cell sizes; two modes at ratio ~1 and $10^{-7}$; **(b)** 350 MPa data with equally distributed 5 cell sizes; two modes at ratio ~1 and $10^{-8}$; **(c)** 350 MPa data with 114 μm cells; two modes at ratio ~1 and $10^{-7}$; **(d)** 350 MPa data with 256 μm cells; two modes at ratio ~1 and $10^{-7}$. Counts of no transition (ratio ~1) decrease with pressure (a-b) and cell size (c-d). Same data also shown in **Figs. 2(a-b)** as Weibull probability plots of a sigmoidal shape.

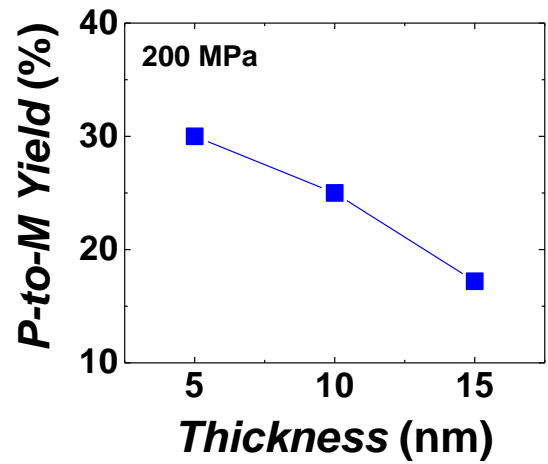

**Figure S3 Relative transition yield under 200 MPa of Pt/HfO$_2$/Ti MIM in different thickenss**. Defined as the accumulative probability of $R$ ratio<$10^{-6}$ in their P-to-M transition Weibull plots, the transition yield decreases as thickness increases, consistent with the picture of percolative network.

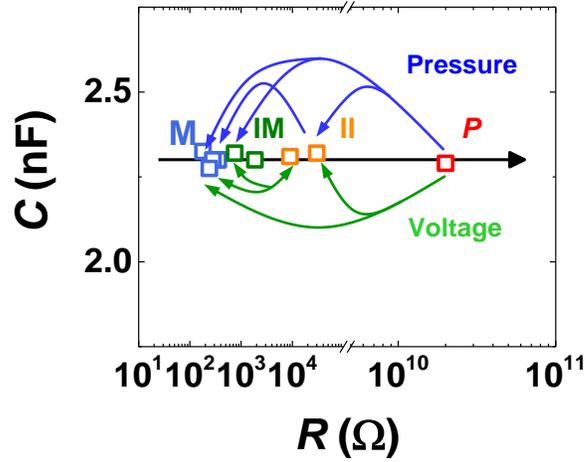

**Figure S4** Same MIM capacitance regardless of resistance in metallic M, intermediate metal IM, intermediate insulating II, and insulating P states, presented in a way similar to **Fig. 1(b)** with arrows indicate various pressure/voltage induced transitions. Capacitance obtained by fitting data of impedance spectroscopy over the frequency range of 10 Hz to 13 MHz.

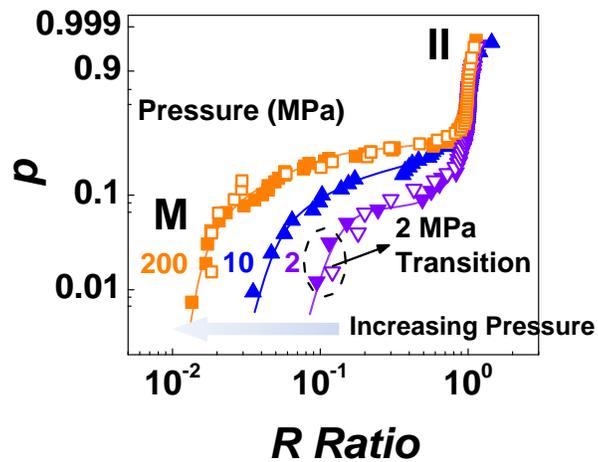

**FIG. S5 II-to-M transition statistics in $HfO_2$ depends on pressure but not how the starting II state was obtained—by either pressure- or voltage-induced transition from P to M state, followed by voltage-transition to II state.** Weibull plots of cumulative pressure-transition probability $p$ in $\ln[\ln[1/(1-p)]]$ scale *vs*. resistance ratio showing higher $p$ at higher pressure. Starting II states that were voltage-transitioned from pressure-induced M are hollow symbols, those from voltage-induced M are solid symbols. Each curve contains 40 preset II cells, equally distributed between 5 cell sizes. Transition at 2 MPa highlighted by dash circle.

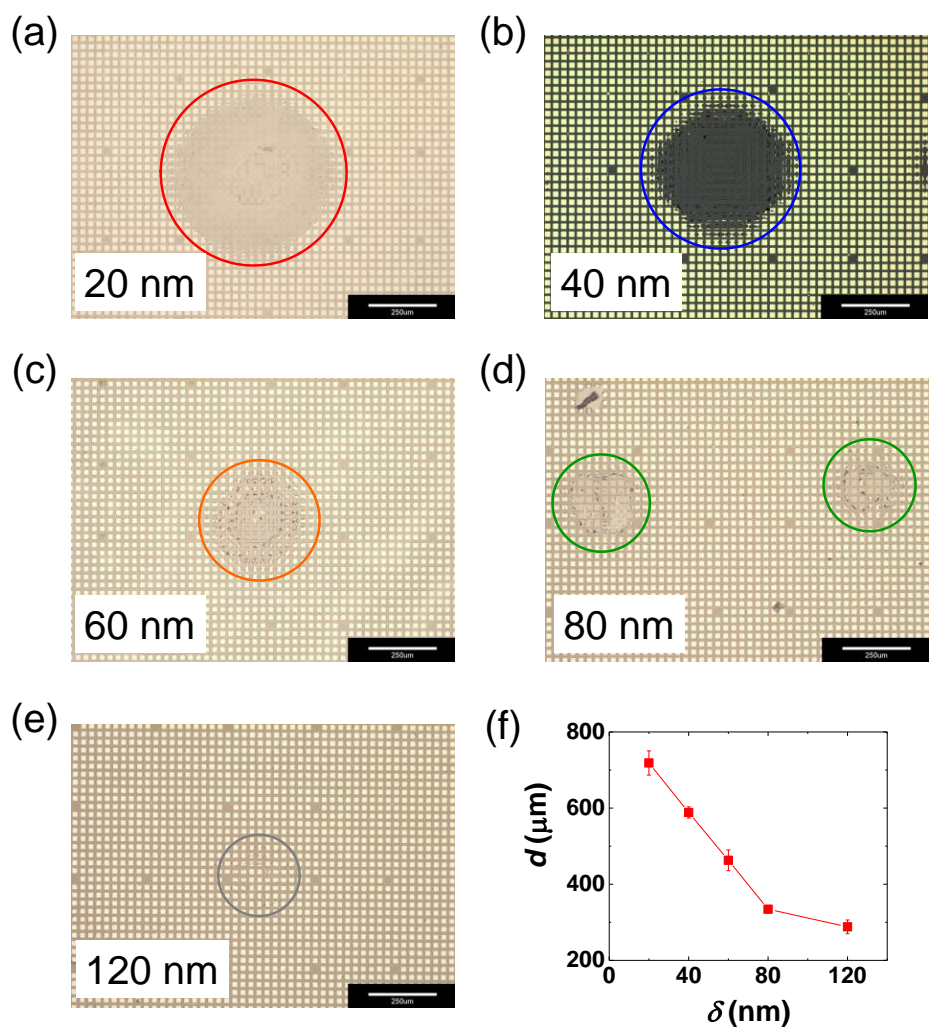

**FIG. S6 Thicker top Pt electrode are not peeled off by electron bunch. (a)-(e)**: Images of the damage zones, defined as regions in which top electrodes have disappeared after one single shot of electron bunch, in arrays of Pt/HfO$_2$/Ti cells with Pt electrode having thickness δ of **(a)** 20 nm, **(b)** 40 nm, **(c)** 60 nm, **(d)** 80 nm, and **(e)** 120 nm where no cell completely lost the electrode. Gold-color squares are undamaged Pt electrodes, while the earth-color regions are the damage zones. Circles outline damage regions with visible deformation, which shrink as Pt electrode becomes thicker as shown in **(f)**. All scale bars on bottom left in **(a-e)** are the same, being 250 μm.

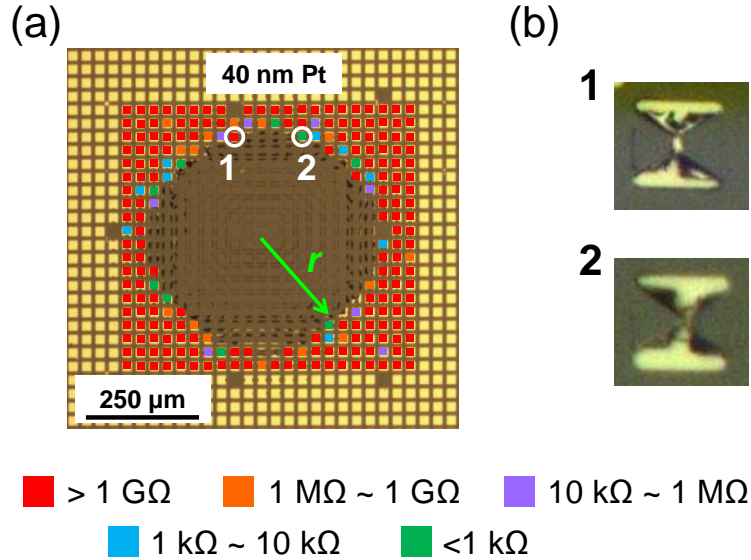

**FIG. S7 Impulse-magnetic-pressure-induced insulator-to-metal transition in $HfO_2$ with thin Pt electrodes.** Resistance map in **(a)** records post-shot resistance values of $Pt/HfO_2/Ti$ MIM cells, originally in P state. Most Pt top electrodes (thickness: 40 nm) were blown off by a single shot of electron bunch making cell resistance unreadable. Cells further away at $r$=250 μm having part of electrodes remaining are readable, but their resistances vary by many orders of magnitude. Cell 1 (red, >1 GΩ) and 2 (green, <1 kΩ), marked by white circles, have vastly different resistances from those of most other cells (in blue, having resistance of 1 kΩ~10 kΩ) at same radial distance. Such feature is a consequence of having top electrode in contact with only a portion of a cell that has undergone percolative transition. The resistance of cell 1 is high because its remaining electrode shown in **(b)** was not in contact with the percolative part of the network. Conversely, the resistance of cell 2 is low because its remaining electrode in **(b)** was in contact with the percolative part of the network.